# Quantum control and measurement of atomic spins in polarization spectroscopy

Ivan H. Deutsch[1] and Poul S. Jessen[2]

Center for Quantum Information and Control (CQuIC)

[1]Dept. of Physics & Astronomy, University of New Mexico, Albuquerque NM 87131, USA

[2]College of Optical Sciences, University of Arizona, Tucson, Arizona 85721, USA

**Abstract:** Quantum control and measurement are two sides of the same coin. To affect a dynamical map, well-designed time-dependent control fields must be applied to the system of interest. To read out the quantum state, information about the system must be transferred to a probe field. We study a particular example of this dual action in the context of quantum control and measurement of atomic spins through the light-shift interaction with an off-resonant optical probe. By introducing an irreducible tensor decomposition, we identify the coupling of the Stokes vector of the light field with moments of the atomic spin state. This shows how polarization spectroscopy can be used for continuous weak measurement of atomic observables that evolve as a function of time. Simultaneously, the state-dependent light shift induced by the probe field can drive nonlinear dynamics of the spin, and can be used to generate arbitrary unitary transformations on the atoms. We revisit the derivation of the master equation in order to give a unified description of spin dynamics in the presence of both nonlinear dynamics and photon scattering. Based on this formalism, we review applications to quantum control, including the design of state-to-state mappings, and quantum-state reconstruction via continuous weak measurement on a dynamically controlled ensemble.

# 1. INTRODUCTION

Atomic spins are robust, coherent, and controllable quantum systems. These features make atomic vapors an ideal platform for applications in precision metrology [1] and quantum information processing [2]. Key tools that enable these applications are the ability to prepare, dynamically evolve, and measure an arbitrary quantum state with high precision in an environment that is sufficiently free from noise and decoherence. As applied to atomic spins, such tools have been developed over the course of decades, with steady advances in coherent spectroscopy, laser cooling and trapping, and quantum optics. In particular, the interface between ensembles of atomic spins and the polarization of optical fields has recently been revived as a rich platform for the exploration of quantum information processing, building on the long history of magneto-optical techniques, originally developed in the context of optical pumping [3]. In the modern laboratory, a wide variety of proof-of-principle experiments have been demonstrated, such as the storage of quantum memory of light in atomic spin vapors [4-11], continuous observation of nonlinear spin oscillations [12, 13], QND measurements of atomic spin ensembles [14-17], the production of spin-squeezed states [18-22], the generation of entanglement between macroscopically separated atomic spin ensembles [23], and quantum-state control/tomography [24-26].

In this article we revisit the interaction between atomic spins and optical probe polarization as a platform for quantum control and measurement. Our goal is to develop a unified and pedagogical treatment, starting from first principles in order to establish the necessary formalism and theoretical description, and then to demonstrate its application in specific protocols that we have carried out in the laboratory. We focus the formalism on alkali atoms, the canonical elements used in most laboratory studies, and particularly $^{133}$Cs, employed in our experiments. Our central concern here is the manipulation of *uncorrelated atoms* in an ensemble of identically prepared and evolving systems. Each atom itself has a rich internal structure, due to the large hyperfine manifolds of magnetic sublevels (e.g., $^{133}$Cs with its valence electron spin $S=1/2$ and nuclear spin $I=7/2$, spans a $d = (2S+1)(2I+1) = 16$ dimensional ground-electronic subspace). The atom-photon coupling acts simultaneously to affect the spin dynamics and as a mechanism to extract quantum information about the spins through polarization spectroscopy. We review our recent work on this subject, including the use of nonlinear spin dynamics to control and prepare arbitrary spin states in a given hyperfine manifold, and to reconstruct the full spin density matrix through weak continuous measurement and control on a single atomic ensemble.

The remainder of the article is organized as follows. In Sec. 2 we establish the fundamental Hamiltonian that governs the tensor nature of the light-shift interaction, expressing it with great utility as a coupling between the probe field's Stokes vector and the corresponding atomic spin variables. The evolution of the Stokes vector on the Poincaré sphere is thus correlated with moments of the atomic spin variables, and polarimetry on the probe can be used as a measurement of the quantum state. The spins' dynamics are generated by a combination of real magnetic fields and light-shift potentials that are decomposed into irreducible tensor effects. A complete (semiclassical) description of the spin dynamics is formulated in terms of a master equation that includes the unavoidable effects of decoherence via photon scattering and optical pumping. In Sec. 3 we apply our formalism to laboratory studies of quantum control and measurement. We first establish the conditions for controllability, and then employ optimal control techniques to design waveforms that generate arbitrary spin states from an initially spin-polarized sample. Monitoring the polarization of the transmitted probe provides a mechanism for continuous weak quantum measurement of the atomic spin. We quantify the conditions under which the measurement strength is sufficiently small that measurement of the probe polarization induces negligible backaction on the atoms. For optically thin samples, this condition is satisfied and continuous weak measurement is seen as monitoring the ensemble average of uncorrelated atoms. We use this to our advantage in implementing a fast and robust protocol for quantum state reconstruction of our large atomic spins. Finally, in Sec. 4 we summarize and give an outlook for future studies.

## 2. THE ATOM-PHOTON TENSOR INTERACTION

### 2.1 The system foundations

A schematic of our quantum control and measurement system is shown in Fig. 1. An ensemble of ~$10^6$ cesium atoms is collected from a magneto-optical-trap (MOT) at a density of ~$10^{10}$ atoms/cm$^{-3}$, cooled in optical molasses and then released into free fall. At this point, the sample is spin polarized via optical pumping, creating a nearly pure spin coherent state of atoms. Control and measurement of the spins are then achieved through the combination of applied magnetic/optical fields and polarimetry analysis of the transmitted probe field, as detailed below.

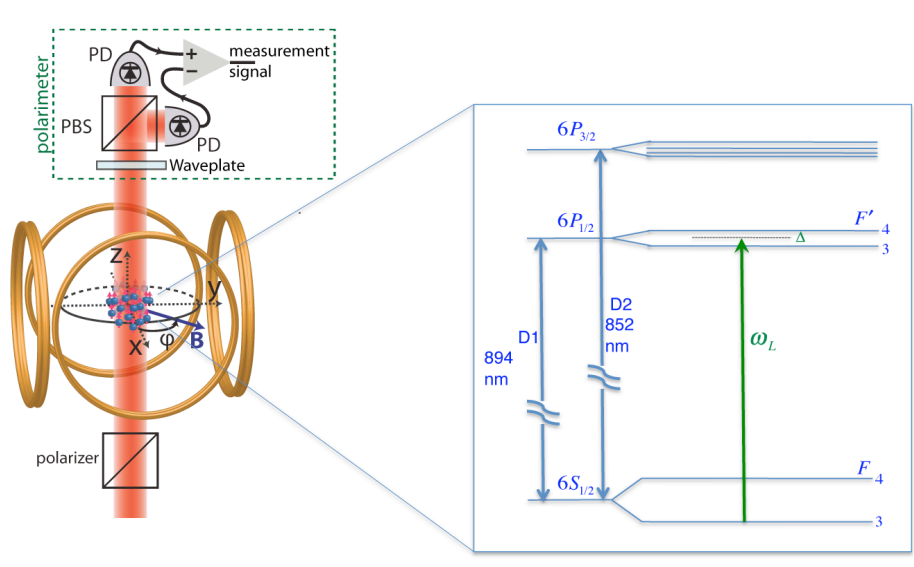

**FIG. 1:** Schematic of our system geometry. A cold gas of atoms is collected from a magneto-optic-trap/optical molasses, and optically pumped to form a nearly pure ensemble of identical spins. The spins are controlled through a combination of light-shift interaction and magnetic fields produced by pairs of Helmholtz coils. A measurement of the spins is performed by polarization analysis of the transmitted probe. The nature of the atom-photon interaction depends strongly on the detuning of the light from resonance. A sketch of the atomic level structure for $^{133}$Cs is shown inset (not to scale). Our probe frequency is typically tuned in-between the two transitions to the $6P_{1/2}$ manifold. We mostly work with atoms in a single ground manifold of given $F$ only; atoms in the other manifold are invisible to the probe field as transitions starting from there are off-resonance by the very large ground hyperfine splitting.

A critical component in the theoretical analysis of this system is the tensor nature of the atom-photon interaction, described in detail by a formalism established decades ago in the context of optical pumping [27]. We lay out here the essential ingredients necessary to describe quantum control and measurement of atomic spins in a more modern context. For a monochromatic light field detuned off resonance, and at sufficiently low intensity that the saturation parameter is small, the atom-light interaction is effectively governed by the dynamics of the magnetic sublevels in the electronic ground-state manifold coupled to the polarization modes of the field. This interaction describes the potential energy of a polarizable particle in an ac-electric field, governed by a Hamiltonian with the same form classically and quantum mechanically,

$$H = -E_i^{(-)} E_j^{(+)} \alpha_{ij}, \tag{1}$$

where $\mathbf{E}^{(\pm)}$ are the positive/negative frequency components of the field and $\alpha_{ij}$ is the atomic dynamic polarizability tensor at frequency $\omega_L$. Quantum mechanically, perturbation theory gives an expression for the polarizability tensor operator,

$$\ddot{\alpha} = -\frac{1}{\hbar} \sum_{g,e} \frac{\mathbf{d}_{ge} \mathbf{d}_{eg}}{\Delta_{eg}}, \tag{2}$$

where $\mathbf{d}_{ge} = \mathbf{d}_{eg}^\dagger = P_g \mathbf{d} P_e$ is the atomic electric dipole operator connecting excited (*e*) and ground (*g*) subspaces via projection operators $P_e$ and $P_g$, and $\Delta_{eg} = \omega_L - \omega_{eg}$ is the detuning of the light field from resonance. The dynamics generated by this interaction affect the field through the spin-dependent index of refraction, and the state of the atoms through the polarization-dependent light-shift.

We restrict our attention to alkali atoms, the species used in most quantum-atom-optics laboratories such as ours; the formalism is easily generalized to other elements. The basic level structure is shown in the inset to Fig. 1, indicating the important regimes of atom-probe coupling. We tune the probe frequency near a strong $nS_{1/2} \to nP_{J'}$ transition, where $J' = 1/2$ (D1 line) or $J' = 3/2$ (D2 line). Here and throughout, primed variables represent excited electronic states, and unprimed variables represent electronic ground states. Associated with each electronic state is a hyperfine multiplet, specified by total angular momentum $\mathbf{F}=\mathbf{I}+\mathbf{J}$, where $\mathbf{I}$ is the nuclear spin. The ground-state hyperfine splitting is typically an order of magnitude larger than that of the excited state. We will consider a detuning on the order of the excited-state hyperfine splitting, a regime for which we obtain rich spin dynamics as discussed below.

The key physical properties of the interaction are seen in a decomposition of the Hamiltonian into its irreducible tensor components [3, 28-31], as detailed in Appendix A. We will assume that the population of interest is restricted to a single ground-state hyperfine manifold with total angular momentum $F$. In that case, according to the Landé projection theorem, the Hamiltonian has an irreducible decomposition of the form,

$$H = -\alpha_0 \left[ C^{(0)} \mathbf{E}^{(-)} \cdot \mathbf{E}^{(+)} + C^{(1)} i \left( \mathbf{E}^{(-)} \times \mathbf{E}^{(+)} \right) \cdot \mathbf{F} + C^{(2)} E_i^{(-)} E_j^{(+)} \left( \frac{1}{2}(F_i F_j + F_j F_i) - \frac{1}{3}\mathbf{F}^2 \delta_{ij} \right) \right], \quad (3)$$

where $\mathbf{F}$ is the angular momentum operator, the tensor coefficients $C^{(K)}$ follow form the Wigner-Eckart theorem as given in Eq. (A13), and the characteristic polarizability $\alpha_0$ depends on the detuning from the excited state. For a fixed excited hyperfine manifold with quantum number $F'$, $\alpha_0 = -\left|\langle P_{J'} \|d\| S_{1/2} \rangle\right|^2 / \hbar\Delta_{F'F} = -\left(3\lambda^3 / 32\pi^3\right)\left(\Gamma/\Delta_{F'F}\right)$, where $\lambda$ is the wavelength of the transition and $\Gamma$ is the spontaneous emission rate (note, cgs units are used throughout). More generally, for the detunings we consider, the Hamiltonian will be a sum of contributions from each state in the excited-state multiplet.

The behavior of the different rank contributions depends on the detuning. Physically, the alkali ground state is a spherically symmetric $l=0$ state. The anisotropic properties of the ac-polarizability arise through the coupling of the orbital angular momentum to the electron and nuclear spins. For detunings that are large compared to the excited-state hyperfine splitting, the optical coupling to the nuclear spin is negligible, and the polarizability depends on of the $s=½$ electron spin. As this spin can only support representations of $SU(2)$, the rank-2 contribution must vanish. Quantitatively, one finds interference between different (indistinguishable) excited state hyperfine channels $F'$. For the D1 and D2 lines, the overall tensor coefficients in the limit of detunings that are large compared to the hyperfine splitting are,

$$\text{D1: } C_{J'F}^{(0)} = \sum_{J'=1/2,F'} C_{J'F'F}^{(0)} = \frac{1}{3}, \quad C_{J'F}^{(1)} = \sum_{J'=1/2,F'} C_{J'F'F}^{(1)} = +\frac{g_F}{3}, \quad C_{J'F}^{(2)} = \sum_{J'=1/2,F'} C_{J'F'F}^{(2)} = 0, \quad (4a)$$

$$\text{D2: } C_{J'F}^{(0)} = \sum_{J'=3/2,F'} C_{J'F'F}^{(0)} = \frac{2}{3}, \quad C_{J'F}^{(1)} = \sum_{J'=3/2,F'} C_{J'F'F}^{(1)} = -\frac{g_F}{3}, \quad C_{J'F}^{(2)} = \sum_{J'=3/2,F'} C_{J'F'F}^{(2)} = 0, \quad (4b)$$

where $g_F$ is the Landé $g$-factor. In the ground-electronic state, in the two hyperfine manifolds $F_{\uparrow(\downarrow)} = I \pm 1/2$, we have $g_{F_\uparrow} = 1/F_\uparrow$ and $g_{F_\downarrow} = -1/(F_\downarrow + 1) = -1/F_\uparrow$ (neglecting the small nuclear magneton). For larger detunings, much greater than the fine-structure splitting, the spin-orbit interaction is negligible and the ac-polarizability loses all spin and polarization dependence, as seen in the vanishing vector contribution to the Hamiltonian through destructive interference, $C_F^{(1)}(D1) + C_F^{(1)}(D2) = 0$.

More precise relations that show the detuning dependencies of the different irreducible tensor contributions follow from a Taylor series expansion. Defining $\Delta_{FF'} = \Delta + \delta_{F'}$, where $\Delta = \Delta_{F'_{max},F}$ and $\delta_{F'}$ is the residual excited-state hyperfine splitting relative to $F'_{max}$, then to order $1/\Delta^2$, the weighted coupling constants in Eq. (3) for the D2 line are

$$\sum_{F'} C_{3/2,F'F_\uparrow}^{(0)} \left( \frac{\Gamma}{\Delta_{FF'}} \right) \approx \frac{2}{3} \left( \frac{\Gamma}{\Delta} \right) - \beta^{(0)} \left( \frac{\Gamma}{\Delta} \right)^2, \quad (5a)$$

$$\sum_{F'} C_{3/2,F'F_\uparrow}^{(1)} \left( \frac{\Gamma}{\Delta_{FF'}} \right) \approx -\frac{1}{3F_\uparrow} \left( \frac{\Gamma}{\Delta} \right) - \beta^{(1)} \left( \frac{\Gamma}{\Delta} \right)^2, \quad (5b)$$

$$\sum_{F'} C_{3/2,F'F_\uparrow}^{(2)} \left( \frac{\Gamma}{\Delta_{FF'}} \right) \approx -\beta^{(2)} \left( \frac{\Gamma}{\Delta} \right)^2, \quad (5c)$$

where the coefficients $\beta^{(i)} = \sum_{F'} (\delta_{F'}/\Gamma) C_{F'}^{(i)}$ are constants that depend on the atomic species, and $\Gamma$ is the excited state linewidth. We see that for detunings that are large compared to the excited-state hyperfine splitting, but small compared to fine structure, the irreducible rank-2 contribution to the tensor interaction is of order $\Gamma/\Delta$ smaller than the vector terms. In fact, this term scales with detuning exactly as the rate of photon scattering. Control and measurement of the atomic spins thus depends on a careful choice of detuning, balancing the requirements of low decoherence with the necessary interactions. We discuss these tradeoffs in detail below.

In the sections below we study the dynamics of the field due to the atomic spin and vice versa. In principle, these are two coupled subsystems, and their joint quantum dynamics should be treated simultaneously. As we will see in Sec. 3.2, for the dilute atomic samples we work with, the quantum entanglement between atoms and photons is negligible, and we can study their evolution separately. In this situation, the quantum state of the light evolves according to the mean field of the atoms and the spins evolve according to the mean field of the light. Strong coupling between the quantum systems will lead to nonclassical correlations beyond the mean-field description, but this regime will not be considered in detail in this article.

**2.2 Field dynamics**

The atom-photon coupling Hamiltonian correlates the dynamics of the atomic spin with the light field. Given the state-dependent tensor polarizability of the atomic sample, this correlation can be used to perform a continuous weak measurement on the atomic spin by monitoring the polarization of a probe beam that traverses the sample, as we will study in Sec. 3.2. Because fundamental fluctuations in the field ultimately limit such a measurement, we include a full quantum mechanical description of the optical modes. We employ a travel-waving quantization scheme discussed in [32], and consider a wave-packet mode of duration $\Delta t$ of a paraxial beam with area $A$. Accounting for the two transverse polarizations (here linear "horizontal" $\mathbf{e}_H$ and "vertical" $\mathbf{e}_V$ polarization), the positive frequency component operator is,

$$\mathbf{E}^{(+)} = \sqrt{\frac{2\pi\hbar\omega}{Ac\Delta t}} \left( \mathbf{e}_H a_H + \mathbf{e}_V a_V \right), \qquad (6)$$

where $a_H, a_V$ are the corresponding photon annihilation operators. Associated with these modes is the Stokes vector $\mathbf{S}$, whose position on the Poincaré sphere represents the polarization state of the field. In quantized form, the corresponding operator is a spin angular momentum via the Schwinger representation for a two-mode oscillator,

$$S_1 = \frac{1}{2}\left(a_H^\dagger a_H - a_V^\dagger a_V\right),\ S_2 = \frac{1}{2}\left(a_H^\dagger a_V + a_V^\dagger a_H\right),\ S_3 = \frac{1}{2i}\left(a_H^\dagger a_V - a_V^\dagger a_H\right), \tag{7}$$

satisfying the *SU*(2) algebra $[S_i, S_j] = i\varepsilon_{ijk} S_k$. We define the total photon number operator, $S_0 = a_H^\dagger a_H + a_V^\dagger a_V$.

The coupling of the Stokes vector to the atomic variables follows from the irreducible tensor decomposition of the Hamiltonian, Eq. (3). From the modal decomposition of the field, Eq. (6), one finds

$$\mathbf{E}^{(-)} \cdot \mathbf{E}^{(+)} = E_{vac}^2 S_0, \tag{8a}$$

$$\mathbf{E}^{(-)} \times \mathbf{E}^{(+)} = 2i E_{vac}^2 S_3 (\mathbf{e}_H \times \mathbf{e}_V), \tag{8b}$$

$$E_i^{(-)} E_j^{(+)} \left(\frac{F_i F_j + F_j F_i}{2}\right) = E_{vac}^2 \left[ S_0\left(\frac{F_H^2 + F_V^2}{2}\right) + S_1(F_H^2 - F_V^2) + S_2(F_H F_V + F_V F_H) \right]. \tag{8c}$$

Plugging these expressions into the Hamiltonian, we arrive at a compact expression for the coupling of the Stokes vector components to the atomic variables,

$$H = \frac{\hbar \chi_0}{\Delta t}\left(A_0 S_0 + A_1 S_1 + A_2 S_2 + A_3 S_3\right), \tag{9}$$

where the operators $A_i$ are atomic observables,

$$A_0 = \left(C^{(0)} - C^{(2)}\left(\frac{3F_k^2 - \mathbf{F}^2}{6}\right)\right),\ A_1 = C^{(2)}\left(\frac{F_H^2 - F_V^2}{2}\right),\ \tag{10a}$$

$$A_2 = C^{(2)}\left(\frac{F_H F_V + F_V F_H}{2}\right),\ A_3 = C^{(1)} F_k, \tag{10b}$$

with $\mathbf{e}_k = \mathbf{e}_H \times \mathbf{e}_V$, the direction of propagation. The dimensionless coupling constant

$$\chi_0 = -\frac{4\pi\omega}{Ac}\alpha_0 = \left(\frac{\sigma_0}{A}\right)\left(\frac{\Gamma}{2\Delta_{F'F}}\right), \qquad (11)$$

is proportional to the fraction of light forward-scattered into the probe mode by one atom, where $\sigma_0 = 3\lambda^2/2\pi$ is the resonant cross section for unit oscillator strength.

When the Hamiltonian is expressed as in Eq. (9), we explicitly see the effects of the atoms on the dynamics of the field. The polarization-independent index of refraction is set by the $S_0$ term and is uninteresting for our purposes here. The remaining terms generate a rotation of the Stokes vector on the Poincaré sphere about an axis and angle depending on the moments of the atomic spin distribution according to a unitary transformation, $U = e^{-iH\Delta t/\hbar} = e^{-i\chi_0 \mathbf{A}\cdot\mathbf{S}}$. Rotation about the **3**-axis precesses the Stokes vector in the equatorial plane of the Poincaré sphere by an amount proportional to the atomic magnetization along the propagation direction – the Faraday effect. Rotation about the **1**-axis or **2**-axis changes the ellipticity of the probe – the signature of birefringence (see Fig. 2). These interactions have the form necessary for QND measurements of the three atomic observables through analysis of each of the three Stokes vector components.

For an ensemble of atoms with density $n_A$ distributed in a volume $V = AL$, we treat the interaction between the probe Stokes vector and the $N_A = n_A A L$ spins in the mean-field approximation, i.e., as a sum over the expectation values of individual atoms, $H_L = \sum_{i=1}^{N_A}\langle H_{AL}^{(i)}\rangle$. Assuming identical states for all spins, the light couples to the atomic ensemble with an effective coupling constant that is $N_A$ times that given in Eq. (11), $N_A \chi_0 = \rho_{OD}\Gamma/(2\Delta_{F'F})$, where $\rho_{OD} = n_A \sigma_o L$ is the "optical density" on resonance. The expected rotation angles of the Stokes vector on the Poincaré sphere are then determined by the mean moments of the atomic spin variables. Following Eq. (10), these rotation angles, $\Theta_i = N_A \chi_0 \langle \hat{A}_i \rangle$, are

$$\Theta_1 = \frac{\rho_{OD}}{4}\langle F_H^2 - F_V^2 \rangle \sum_{F'} C_{J'F'F}^{(2)} \frac{\Gamma}{\Delta_{FF'}}, \qquad (12a)$$

$$\Theta_2 = \frac{\rho_{OD}}{4}\langle F_H F_V + F_V F_H \rangle \sum_{F'} C_{J'F'F}^{(2)} \frac{\Gamma}{\Delta_{FF'}}, \qquad (12b)$$

$$\Theta_3 = \frac{\rho_{OD}}{2}\langle F_k\rangle \sum_{F'} C^{(1)}_{J'F'F}\frac{\Gamma}{\Delta_{FF'}}. \qquad (12c)$$

In the limit where $\Theta_i \ll 1$, the field's Stokes vector is transformed according to

$$\mathbf{S}_{out} = \mathbf{S}_{in} + \vec{\Theta}\times\mathbf{S}_{in}. \qquad (13)$$

Maximum sensitivity in the measurement is achieved when the polarimeter analyzes the output along a direction on the Poincaré sphere (defined here as $\mathbf{e}_{out}$) orthogonal to the input. In this case, the signal is approximately proportional to $\mathbf{e}_{out}\cdot\mathbf{S}_{out} = (\mathbf{S}_{in}\times\mathbf{e}_{out})\cdot\vec{\Theta}$. For example, taking the input polarization along the $\mathbf{e}_H$-direction, $\mathbf{S}_{in} = \mathbf{e}_1$, an output analyzer along the linear diagonal directions ($\pm 45°$), $\mathbf{e}_{out} = \mathbf{e}_2$, performs an ensemble measurement of $\langle F_k\rangle$ via Faraday rotation, $\mathbf{e}_{out}\cdot\mathbf{S}_{out} \sim \Theta_3$, whereas an output analyzer in the circular basis, $\mathbf{e}_{out} = \mathbf{e}_3$, performs a measurement of the second order atomic moment $\langle F_H F_V + F_V F_H \rangle$ due to birefringence in the sample, $\mathbf{e}_{out} \cdot \mathbf{S}_{out} \sim \Theta_2$. Such measurements can be used to continuously observe the dynamics of the atomic spin. Examples of continuous measurement of this kind will be discussed in Sec. 3.2.

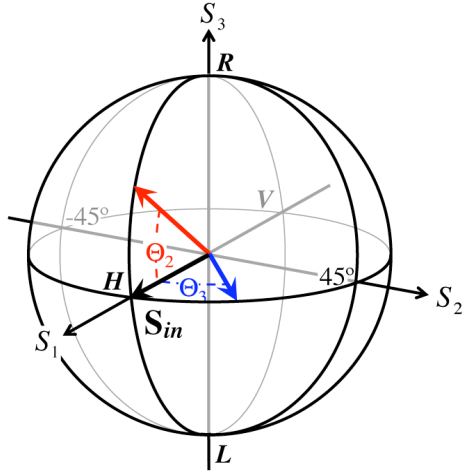

**FIG. 2:** Transformation of the Stokes vector on the Poincaré sphere. For an initially linearly polarized field along the $S_1$ direction, a rotation on the sphere about $S_3$ by an angle $\Theta_3 \propto \langle F_k\rangle$ corresponds to a physical rotation of the polarization vector by $\Theta_3/2$, proportional the atomic magnetization (the Faraday effect). A rotation about $S_2$ by an angle $\Theta_2 \propto \langle F_H F_V + F_V F_H\rangle$ corresponds to a change in the ellipticity of the light (birefringence). For small angles of rotations, as occur in atomic vapors at large probe detunings, the rotation angles and their correlation to associated atomic observables appear as local displacements on the sphere. The Faraday effect is thus measured as a displacement along $S_2$ and birefringence as a displacement along $S_3$. These polarization changes can be detected as an intensity balance in the photo-detectors of an appropriately configured polarimeter (see Fig. 1).

**2.3 Spin dynamics**

The atom-probe coupling Hamiltonian depends on the atomic spin degrees of freedom through the tensor polarizability, and thus have the potential to drive non-trivial dynamics of the atomic spins. This evolution complements the transformation of the probe Stokes vector described in the previous section. We restrict our attention to probe fields described by a coherent state $|\mathbf{E}_0\rangle$, defined by $\mathbf{E}^{(+)}|\mathbf{E}_0\rangle = (\mathbf{E}_0/2)|\mathbf{E}_0\rangle$, where $\mathbf{E}_0$ is the classical complex amplitude for a probe polarization $\vec{\varepsilon}$, so that the expectation value of the field is $\mathbf{E} = \text{Re}(E_0 \vec{\varepsilon} e^{-i\omega_L t})$. With a coherent state input and no measurement backaction the atoms remain uncorrelated, evolving independently according to a light-shift Hamiltonian set by the probe mean-field,

$$H_A = \langle \mathbf{E}_0 | H_{AL} | \mathbf{E}_0 \rangle = \sum_{F'} V_0 \left[ C^{(0)}_{J'F'F} |\vec{\varepsilon}|^2 + iC^{(1)}_{J'F'F} (\vec{\varepsilon}^* \times \vec{\varepsilon}) \cdot \mathbf{F} + C^{(2)}_{J'F'F} \left( |\vec{\varepsilon} \cdot \mathbf{F}|^2 - \frac{1}{3} \mathbf{F}^2 |\vec{\varepsilon}|^2 \right) \right], \quad (14)$$

where $|\vec{\varepsilon} \cdot \mathbf{F}|^2 = \frac{1}{2}\left[ (\vec{\varepsilon} \cdot \mathbf{F})^\dagger (\vec{\varepsilon} \cdot \mathbf{F}) + (\vec{\varepsilon} \cdot \mathbf{F})(\vec{\varepsilon} \cdot \mathbf{F})^\dagger \right]$. The overall scale,

$$V_0 = -\frac{1}{4}\alpha_0 |E_0|^2 = \frac{\hbar \Omega^2}{4\Delta_{FF'}} = \left( \frac{\hbar \Gamma}{8} \frac{I}{I_{sat}} \right) \frac{\Gamma}{\Delta_{FF'}}, \quad (15)$$

is the ac-Stark shift (light shift) associated with a field of intensity $I$ acting on a transition with unit oscillator strength and saturation intensity $I_{sat}$, detuned by $\Delta_{F'F}$ from a hyperfine resonance, where $\Omega = \langle nP_{J'} \|d\| nS_{1/2} \rangle E_0 / \hbar$ is the characteristic Rabi frequency of the D1 or D2 line.

The decomposition of the light-shift interaction into its irreducible tensor components reveals physically distinct effects. The rank-0 component produces an equal energy level shift for all sublevels within a ground hyperfine manifold, depending only on the total intensity of the field, and independent of its polarization $\vec{\varepsilon}$. For systems restricted to a given manifold $F$, this term does not drive atomic dynamics. The rank-1 component is a Zeeman-like interaction, $H^{(1)} = \mathbf{B}_{fict} \cdot \mathbf{F}$, with a fictitious magnetic field proportional to $i(\vec{\varepsilon}^* \times \vec{\varepsilon})$, and thus depends on the ellipticity of the probe polarization (photon spin). This component generates rotations of the atomic spin about $\mathbf{B}_{fict}$. The rank-2 component contains a nonlinear light-shift proportional to $|\vec{\varepsilon} \cdot \mathbf{F}|^2$, generating dynamics beyond $SU(2)$ rotations.

The relative strength of these different contributions depends on the polarization of the light and the detuning from resonance. For detunings that are small compared to the excited state hyperfine splitting, the rank-2 and rank-1 contributions are of the same order for a generic elliptical polarization. For much larger detunings, the rank-1 dominates over the rank-2 contribution to the light shift. The exception is for linear polarization, in which case the fictitious magnetic field vanishes and the light shift within a hyperfine manifold takes the form

$$H_A = \sum_{F'} V_0 \, C^{(2)}_{J'F'F} (\vec{\varepsilon} \cdot \mathbf{F})^2 \equiv \beta^{(2)} \hbar \gamma_s (\vec{\varepsilon} \cdot \mathbf{F})^2, \tag{16}$$

where we have omitted constant terms that are independent of the spin projection. In the second expression for this nonlinear light shift Hamiltonian we have explicitly factored out the unit-oscillator-strength photon scattering rate, $\gamma_s = \Omega^2 \Gamma/(4\Delta^2)$. We do this for two reasons. First, our ability to drive control coherent spin dynamics is limited by decoherence arising from photon scattering. The coefficient $\beta^{(2)}$ thus roughly determines how many nonlinear oscillations can occur in a photon scattering time. Secondly, though in general $\beta^{(2)}$ depends on detuning, for detunings large compared to the excited state hyperfine splitting, this coefficient becomes a constant, according to Eq. (5c). This reminds us that the rank-2 light shift can never be made arbitrarily large compared to photon scattering, but in fact scales with detuning in exactly as a scattering rate in the large detuning limit. The ability to implement nonlinear spin dynamics depends on achieving a favorable balance between the two, as discussed in further detail in Sec. 3.

These considerations show that a complete description of spin-control via the nonlinear light shift must include the effects of photon scattering in a quantitative manner. In the low saturation limit, one can obtain a master equation solely for the ground electronic manifold via adiabatic elimination, as outlined in Appendix B. The master equation for a single ground-state manifold $F$ is

$$\frac{d\rho}{dt} = -\frac{i}{\hbar}\left(H_A^{eff}\rho - \rho H_A^{eff\dagger}\right) + \Gamma\sum_q W_q \rho W_q^\dagger. \tag{17}$$

Here, the non-Hermitian effective atomic light-shift Hamiltonian is obtained by substituting $\Delta \to \Delta + i\Gamma/2$ in the definition of the polarizability in Eq. (2) and taking the fields to be classical under the coherent-state assumption as in Eq. (14),

$$H_A^{eff} = -\frac{1}{4}E_i^* E_j \alpha'_{ij}, \quad \ddot{\alpha}' = -\frac{1}{\hbar}\sum_{F'}\frac{\mathbf{d}_{FF'}\mathbf{d}_{F'F}}{\Delta_{F'F} + i\Gamma/2}. \tag{18}$$

In the off-resonance limit, the effective Hamiltonian is well approximated through the addition of an imaginary part to the light-shift amplitude in Eq. (15), $V_0 \to V_0 - i\hbar\gamma_s/2$. The sum term in the master equation Eq. (17) accounts for optical pumping that returns excitation back to the hyperfine manifold of interest. Each "jump process" corresponds to a cycle of absorption of a probe photon of polarization $\vec{\varepsilon}$, excitation from the ground-manifold $F$ to some excited manifold, $F'$, emission of a spontaneous photon with polarization $\mathbf{e}_q$, and decay back to the original manifold. As shown in Appendix B, adiabatic elimination gives the jump operators as,

$$W_q = \sum_{F'}\frac{\Omega/2}{\Delta_{F'F} + i\Gamma/2}\left(\mathbf{e}_q^* \cdot \mathbf{D}_{FF'}\right)\left(\vec{\varepsilon}_L \cdot \mathbf{D}_{F'F}^\dagger\right). \tag{19}$$

Here $\mathbf{e}_q \cdot \mathbf{D}_{F'F}^\dagger = \sum_{M,M'} o_{JF}^{J'F'} \langle F'M'|FM;1q\rangle \, |F'M'\rangle\langle FM|$ are the dimensionless raising operators from ground to excited sublevels, accounting for the Clebsch-Gordan coefficients for the different dipole transitions and the relative oscillator strengths for the transitions $S_{1/2,F} \leftrightarrow P_{J',F'}$, $o_{JF}^{J'F'}$ (see Eq. A2). In the rate equation limit, it follows that the optical pumping transition rate from $Fm_1 \to Fm_2$ is,

$$\gamma_{FM_1 \to FM_2} = \sum_q |\langle FM_2|W_q|FM_1\rangle|^2 = \frac{\Omega^2 \Gamma}{4} \sum_q \left| \sum_{F'} \frac{\langle FM_2|(\mathbf{e}_q^* \cdot \mathbf{D}_{FF'})(\vec{\varepsilon} \cdot \mathbf{D}_{F'F}^\dagger)|FM_1\rangle}{\Delta_{F'F} + i\Gamma/2} \right|^2, \quad (20)$$

in agreement with the Kramers-Heisenberg formula [33]. This equation explicitly shows the interference between different scattering paths through virtual excitation to the excited states $F'$. Note that the master equation in Eq. (17) is not trace-preserving, as population can be optically pumped to the other ground hyperfine manifold. The large ground hyperfine splitting ensures both that this population is invisible to the probe, and that it is unlikely to be optically repumped back into the manifold of interest. A more complete description, including all hyperfine sublevels, can be found in Appendix B.

For detunings much larger than the excited state hyperfine splitting, processes associated with the irreducible rank-2 tensor interfere destructively. In addition to a reduction of the light-shift interaction to only scalar and vector terms, optical pumping transitions with $\Delta M = \pm 2$ vanish. Quantitatively, using the decomposition of the tensor into its irreducible components Eq. (A12), the jump operators can be expressed as,

$$\mathbf{e}_q^* \cdot (\mathbf{D}_{FF'} \mathbf{D}_{F'F}^\dagger) \cdot \vec{\varepsilon}_L = C_{J'F'F}^{(0)} \mathbf{e}_q^* \cdot \vec{\varepsilon}_L + i C_{J'F'F}^{(1)} (\mathbf{e}_q^* \times \vec{\varepsilon}_L) \cdot \mathbf{F} + $$
$$+ C_{J'F'F}^{(2)} \left[ \frac{(\mathbf{e}_q^* \cdot \mathbf{F})(\vec{\varepsilon}_L \cdot \mathbf{F}) + (\vec{\varepsilon}_L \cdot \mathbf{F})(\mathbf{e}_q^* \cdot \mathbf{F})}{2} - \frac{1}{3} |\mathbf{e}_q^* \cdot \vec{\varepsilon}_L|^2 \mathbf{F}^2 \right]. \quad (21)$$

When the detuning is approximately independent of $F'$ it can be factored out of the sum, and the spontaneous "feeding" terms in the master equation become,

$$\left.\frac{d\rho}{dt}\right|_{feed} = \gamma_s \sum_q \tilde{W}_q \rho \tilde{W}_q^\dagger, \quad (22a)$$

$$\tilde{W}_q = \sum_{F'} \mathbf{e}_q^* \cdot \left(\mathbf{D}_{FF'} \mathbf{D}_{F'F}^\dagger\right) \cdot \vec{\varepsilon}_L = C_{J'F}^{(0)} \mathbf{e}_q^* \cdot \vec{\varepsilon}_L + i C_{J'F}^{(1)} \left(\mathbf{e}_q^* \times \vec{\varepsilon}_L\right) \cdot \mathbf{F}, \quad (22b)$$

with the coefficients $C_{J'F}^{(K)}$ defined in Eq. (4).

As an example, consider an atom driven on the D2 line by a probe that is linearly polarized and detuned far from the excited state hyperfine multiplet compared to the hyperfine splitting. The effective atomic Hamiltonian is approximately given as, $H_A^{eff} = H_A - i\hbar(\gamma_s/3)\hat{1}$, and the jump operators become

$$\tilde{W}_0 = \frac{2}{3}\hat{1}, \quad \tilde{W}_{\pm 1} = \frac{g_F}{3\sqrt{2}} F_\mp, \quad (23)$$

where $F_\pm = F_x \pm i F_y$ are the angular momentum raising and lowering operators. The master equation then takes the form,

$$\frac{d\rho}{dt} = -\frac{i}{\hbar}[H_A, \rho] - \frac{2}{9}\gamma_s \rho + \frac{g_F^2}{18}\gamma_s \left(F_+ \rho F_- + F_- \rho F_+\right). \quad (24)$$

As a comparison, consider a hypothetical atom with zero nuclear spin. In that case, given linear polarization, $H_A \propto \hat{1}$, $g_F = g_s = 2$, and thus,

$$\frac{d\rho}{dt} = -\frac{2}{9}\gamma_s \rho + \frac{2}{9}\gamma_s \left(\sigma_+ \rho \sigma_- + \sigma_- \rho \sigma_+\right). \quad (25)$$

This is a trace preserving map describing optical pumping in the two-level system, familiar from, e.g., the theory of Sisyphus cooling [34].

As an example of nontrivial spin dynamics, we consider Larmor precession of a spin **F** in a static magnetic field, in the presence of an *x*-linearly polarized probe field that gives rise to both a tensor light shift and photon scattering. Choosing the static magnetic field along the *x*-direction, the master equation is given by Eq. (17), with the effective

Hamiltonian now accounting for both magnetic and optical interactions, taking the form $H_A^{eff} = \hbar\Omega_{Lar}F_x + \hbar\gamma_s(\beta^{(2)} - i/2)F_x^2$, where $\Omega_{Lar} = g_F\mu_B B/\hbar$ is the Larmor frequency, and $\beta^{(2)}$ is defined in Eq. (16). For concreteness, we consider an experiment with $^{133}$Cs atoms initially prepared in the $F = 3$ manifold with maximum projection along $y$. We set the Larmor frequency to $\Omega_{Lar}/2\pi = 17.5$ kHz, and tune the probe field ~524 MHz to the red of the $6S_{1/2}(F=3) \to 6P_{1/2}(F'=4)$ transition, approximately halfway between the two components of the D1 line, where the tensor light shift is greatest and the nonlinear coefficient $\beta^{(2)} \approx 8.2$. The probe intensity is chosen so the average time between photon scattering events is $\gamma_s^{-1} \approx 2.5$ ms, and the period of nonlinear oscillation is $\left(\beta^{(2)}\gamma_s/2\pi\right)^{-1} \approx 1.9$ ms. Figure 3 shows the corresponding evolution of $\langle F_z(t)\rangle$, consisting of rapid oscillations at the Larmor frequency, a series of collapse and revivals on the expected timescale for nonlinear oscillation, and an overall signal decay on a timescale roughly equal to $\gamma_s^{-1}$. In this particular example, the overall decay occurs in part due to photon scattering and optical pumping, and in part due to variations in probe intensity and accompanying nonlinear strength $\propto \gamma_s$ across the spin ensemble. The latter leads to a spread in the timescale for nonlinear oscillation, and thus a smearing and reduction in the amplitude of the later revivals. Nevertheless, for the parameters chosen here we see at least six distinct revivals, demonstrating that a considerable degree of nonlinear evolution can be achieved before the system decoheres or dephases. Note also the close agreement between the prediction of a full theoretical model based on the master equation (Eq. 17) and including a statistical average over the nonlinear strengths (Fig. 3a), and the experimentally measured signal (Fig. 3b). This underscores the considerable precision with which the system can be characterized in the laboratory, and modeled according to the formalism presented here. These capabilities form the foundation for applications in quantum control and measurement, as discussed in more detail in the following section.

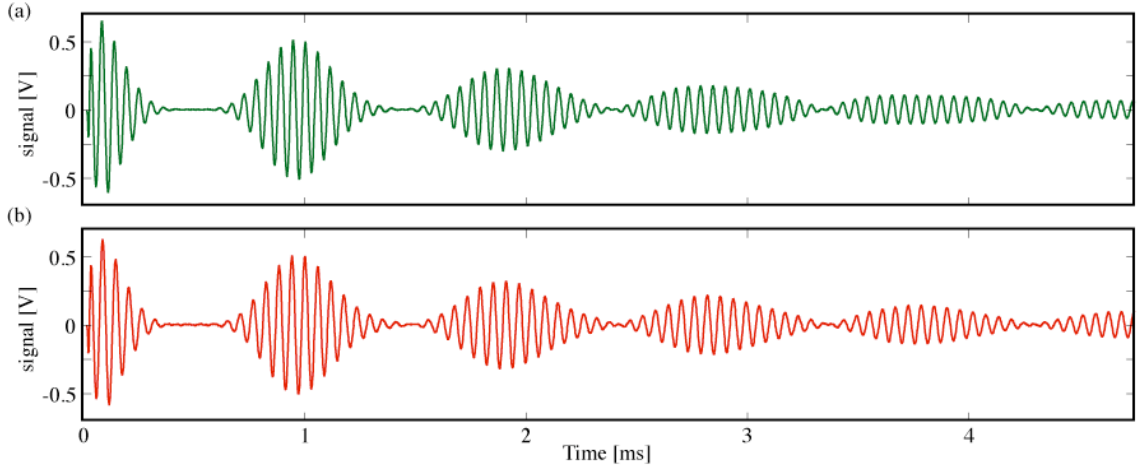

**Fig. 3:** Larmor precession of a hyperfine spin in the presence of a static magnetic field and a linearly polarized probe field. (a) Theoretical model based on the master equation (17) and including an average over a distribution of nonlinear strengths, as described in the text. (b) Experimental data measured via polarization spectroscopy. The nonlinearity gives rise to collapse and revivals of the Larmor precession signal, while photon scattering and variations in the nonlinearity both contribute to overall damping at later times. Detailed parameter values are given in the text.

## 3. APPLICATION TO QUANTUM CONTROL AND MEASUREMENT

Quantum control and measurement are typically considered separate tasks, but when both are accomplished by coupling the quantum system of interest to ancillary fields they become flip sides of the same coin. In quantum control scenario, the goal is to affect a dynamical map on the quantum system through the application of one or more control fields, while the goal of quantum measurement is to transfer quantum information about the system to one or more probe fields, which are subsequently read out as macroscopic classical signals. In our case the coupling between atomic spins and the polarization of a single probe field is central to both, simultaneously driving spin dynamics according to Hamiltonian Eq. (14) and rotations of the Stokes vector on the Poincaré sphere according to Eq. (9). In this section we describe in detail some of the protocols we have designed and implemented for control and measurement in this system.

### 3.1 Quantum Control

In closed-system dynamics, a quantum system is said to be *controllable* if, given the available Hamiltonian interactions, there exists a possible route to implement an arbitrary unitary map on the Hilbert space of interest. Group theory provides the natural framework in which to study quantum controllability [35]. For a *d*-dimensional Hilbert space, the goal is to implement an arbitrary map $U$ in the group $SU(d)$. At our disposal is a time-dependent a Hamiltonian of the form,

$$H(t) = \sum_{j=1}^{j_{max}} \lambda^{(j)}(t) h_j, \qquad (26)$$

where $\{\lambda^{(j)}(t)\}$ are c-number "control waveforms" that are functions of the applied classical fields, and $\{h_j\}$ are a set of available time independent Hamiltonians.

All unitary maps generated by the Hamiltonian are functionals of the control fields according to the time-ordered product,

$$U_t[\lambda^{(1)}, \ldots, \lambda^{(j_{max})}] = T\left\{\exp\left(-\frac{i}{\hbar} \sum_j \int_0^t dt' \lambda^{(j)}(t') h_j\right)\right\}. \qquad (27)$$

In general, the infinitesimal generators of these maps form a Lie algebra, defined as the span of linear combinations of $\{h_j\}$ and any multiple commutators of members of the set. A system with this time-dependent Hamiltonian, acting on a finite dimensional Hilbert space of dimension $d$ is then said to be controllable if and only if $\{h_j\}$ is a generating set for the full Lie algebra of interest, $su(d)$ [35].

In the context of the atomic system at hand, controllability of a spin in a hyperfine manifold with total angular momentum $F$ requires a set of Hamiltonian interactions that generate the Lie group $SU(2F+1)$. For a linear Zeeman interaction $H_Z(t) \sim \mathbf{B}(t) \cdot \mathbf{F}$, the set of scaled control Hamiltonians, $\{F_x, F_y, F_z\}$, form a basis for the Lie algebra $su(2)$. Since the group $SU(2)$ consists solely of geometric rotations, this interaction leads to full controllability only for $F=1/2$. For $F>1/2$ one requires an interaction that is nonlinear in some component of the spin, such as the Hamiltonian arising from the irreducible rank-2 light shift in Eq. (16). Consider, for example, the light-shift interaction induced by a probe field that is linearly polarized along the *x*-direction. Ignoring constant terms within the hyperfine subspace, the light-shift Hamiltonian reduces to Eq. (16) and takes the form $H_{LS} = \beta^{(2)}\hbar\gamma_s F_x^2$. Such an interaction is known to produce squeezed states [36] and so-called "cat-states" in $F>1/2$ spin systems [37], and must therefore be able to generate maps that are more general than geometrical rotations. To achieve full controllability, one must combine this nonlinear interaction with other noncommuting Hamiltonians. A minimal generating set is $\{F_x, F_y, F_x^2\}$. This can be seen as follows. Multiple commutators of the linear set $\{F_x, F_y\}$ close to the finite set $\{F_x, F_y, F_z\}$, and thus generate $su(2)$. Commutators with the nonlinear terms, however, produce new generators. For example, $[F_x^2, F_z] = -i(F_x F_y + F_y F_x)$ is an addition to the set. Multiple commutators will eventually span all polynomials in the components of **F**, and thus the entire Lie algebra of $su(2F+1)$, for an arbitrary $F$.

We can implement this minimal generating set based on the Zeeman interaction between an atomic spin and a constant magnitude magnetic field rotating in the *x-y* plane, combined with a constant nonlinear light shift from an *x*-polarized probe field. The total control Hamiltonian then takes the form,

$$H(t) = \hbar\Omega_{Lar}\left(\cos\theta(t)F_x + \sin\theta(t)F_y\right) + \beta^{(2)}\hbar\gamma_s F_x^2, \tag{28}$$

where $\theta(t)$ is the time-dependent angle of the magnetic field with respect to the *x*-axis.

Given the Hamiltonian in Eq. (28), a unitary map on the spin is determined by the time-dependent direction of the magnetic field, set by $\theta(t)$. While controllability guarantees that such a function exists for any unitary matrix of interest, in general there are no constructive algorithms for finding it. Instead, one must employ the methods of "optimal control" and perform a numerical search for the best approximation to *U* by maximizing an objective function with respect to the control waveform. The computational complexity of such an optimization depends on the quantum control task at hand. In a series of papers, Rabitz and coworkers explored the structure of the "control landscape" [38-41], i.e., the hypersurface representing the objective as a function of the control waveform. The landscape for a general unitary map is not favorable, and empirical studies show that the numerical effort required to find optimal solutions grows exponentially with the dimension of the Hilbert space [41].

A simple but important special control task is state-to-state mapping, or state-preparation, i.e., finding a control waveform that maps a known fiducial state $|\psi_0\rangle$ to an arbitrary desired target state $|\psi_t\rangle$. Imposing only the requirement $|\psi_t\rangle = U|\psi_0\rangle$ is equivalent to specifying just one column of a unitary matrix, with no constraint whatsoever on the action of $U$ on the orthogonal complement to $|\psi_0\rangle$. In this situation there is no unique solution, and the added freedom in designing $U$ makes the topology of the control landscape much more amenable to numerical optimization. In an important theorem, Rabitz *et al.* proved that for perfect Hamiltonian evolution with an arbitrarily long duration, all local optima of the objective function are also global optima [38], and each optimum is located on a high-dimensional manifold that slopes gently towards it [39]. This favorable landscape greatly simplifies the search for control waveforms, guaranteeing that a simple gradient ascent from a random starting point will end in a global optimum. Note that the theorem is strictly true only in the ideal case, with no decoherence, no errors in the control Hamiltonian, and arbitrarily long control time. Nevertheless, we have found that control waveforms determined for these ideal conditions serve as excellent starting points for further optimization that take into account experimental imperfections.

Numerical optimization of the control waveform requires that it be specified by a finite number of parameters. To do this we pick some fixed control time $T$, short compared to the decoherence time but long enough to reach an arbitrary state, coarse-grain by specifying the angles at discrete times, $\theta_i = \theta(t_i)$, and interpolate between them in a manner consistent with experimental constraints on the bandwidths and slew rates of the control fields. Propagating the state according to pure Hamiltonian evolution allows us to determine the final state, $|\psi(T)\rangle = U(\vec{\theta})|\psi_0\rangle$, as a function of the vector $\vec{\theta}$ containing the coarse-grained values $\theta_i$. Having set up the problem in this way, we can optimize the fidelity of the prepared state relative to the target state, $\mathcal{F}(\vec{\theta}) = |\langle\psi_t|U(\vec{\theta})|\psi_0\rangle|^2$. Given a random initial seed vector, $\vec{\theta}^{(0)}$, a search through a gradient ascent, $\vec{\theta}^{(n+1)} = \vec{\theta}^{(n)} + \varepsilon \vec{\nabla}\mathcal{F}(\vec{\theta})$, will quickly converge on a local optimum. Empirically, we find that the finite control time available for our system is not a significant limitation, and that we can reach a typical fidelity >99% from an arbitrary seed. Having found a set of candidate control waveforms, it is usually desirable to perform a second round of optimization that accounts for decoherence from photon scattering, as well as statistical variations in the control waveform across our atomic ensemble. A simple way to do this is to check the candidate waveforms one by one, plugging the control fields into the master equation, Eq. (17), and integrating to find the corresponding final state $\rho(\vec{\theta})$. The optimal choice is then the waveform that maximizes the fidelity $\mathcal{F}(\vec{\theta}) = \langle \psi_t | \rho(\vec{\theta}) | \psi_t \rangle$.

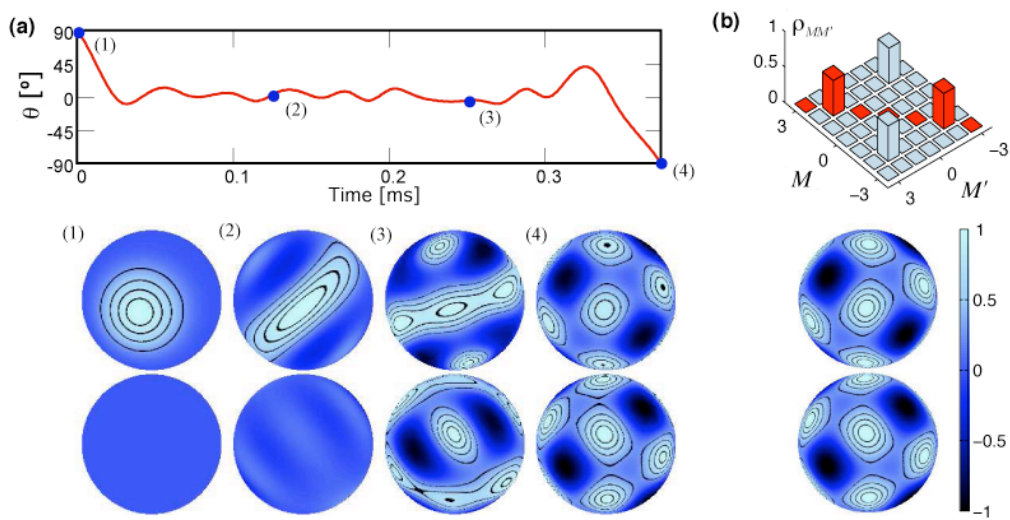

**FIG 4:** Quantum control of an atomic spin. (a) Example of a control waveform $\theta(t)$ specifying the direction of a constant-magnitude magnetic field in the *x-y* plane. (1-4) Wigner functions calculated at four stages during the control sequence. Both Bloch hemispheres are shown. The final result is close to the target state $|\psi_t\rangle \propto |M=2\rangle + |M=-2\rangle$ (b) Density matrix (absolute values) and Wigner function for $|\psi_t\rangle$. From [26].

We have tested the above procedure in experiments with ensembles of laser cooled and optically pumped Cs atoms, as described in detail in [26]. As discussed in Sec. 2.3, Cs is a good choice for such work because of the large hyperfine splitting of the $6P_{1/2}$ excited state manifold. By tuning the probe frequency approximately halfway between the transitions to the $6P_{1/2}(F'=3)$ and $6P_{1/2}(F'=4)$ states, we obtain a nonlinear coefficient $\beta^{(2)} \approx 8.2$, the largest possible for any of the alkalis. Since $\beta^{(2)}$ is a measure of the relative rates of nonlinear evolution and decoherence from photon scattering, this value suggests that quantum state mapping is possible but that a non-negligible infidelity is likely to result from photon scattering. Detailed simulations for the same parameters as used in Fig. 3 show that fidelities in excess of 90% are possible if there are no errors in the control fields. Figure 4 shows an example of a control waveform, designed to map an initial state $|F=3, M=3\rangle$ in the $F=3$ ground state manifold to a nontrivial superposition of magnetic sublevels. Also shown are theoretical predictions for the evolving state, as expressed on the Bloch sphere by a spin Wigner-function representation. For clarity the states are shown in a rotating frame to transform away overall rotations caused by the control magnetic field. We see that the effect of the nonlinear light-shift is to squeeze the nitial spin coherent state, as expected for an $F_x^2$ interaction. Eventually, the Wigner function is elongated to the point where it wraps around the Bloch sphere and nonclassical interference is exhibited. The role of the time-dependent magnetic field is to move and rotate the Wigner function on the Bloch sphere, thereby modifying the squeezing and interference so that, at the end of the waveform, the evolved state comes as close as possible to the target state. Note that the cumulative rotation on the Bloch sphere is quite large, and that a small over- or under-rotation can degrade the fidelity considerably. Figure 5 shows examples of three target states, along with the states produced in the laboratory. The latter were determined by the quantum state reconstruction procedure outlined in the next section. For these examples the observed fidelities are quite high; in the course of more than 100 trials producing 21 different states, we observe typical fidelities in the range 70-90%. Factoring out small under- and over-rotations and taking into account the ~90% fidelity of our quantum state reconstruction puts most of these fidelities in the 80-90% range, in good agreement with the ~90% predicted by the theoretical model used to design the control waveforms in the first place.

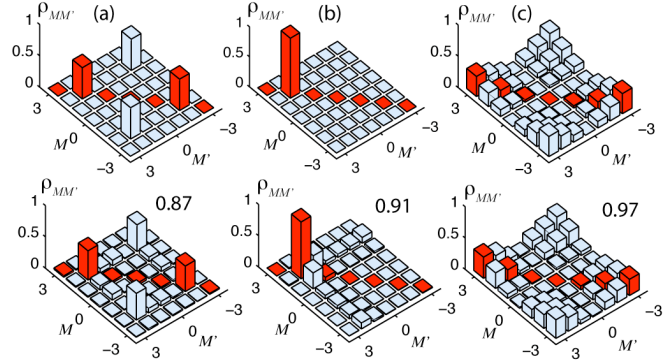

**Fig. 5:** Examples of target and measured density matrices (absolute values). The target states are (a) $(|M_z = 2\rangle + |M_z = -2\rangle)/\sqrt{2}$, (b) $|M_x = 2\rangle$, and (c) $\sum_{M_y} M_y |M_y\rangle$. The fidelity between target and measured states is indicated in each case. From [26].

## 3.2 Measurement

### 3.2.1 Quantum measurement backaction

Kastler first proposed the use of off-resonant light to probe the spin state of atoms [42], and the idea was subsequently demonstrated in experiments by Manuel and Cohen-Tannoudji [43]. In the modern quantum optics context, the light shift interaction between atomic spins and a probe field provides a realization of quantum measurement in the classic von Neumann paradigm – the quantum system (spins) is coupled to a quantum "meter" (probe polarization) for some time, and then the meter is read out by polarization analysis. In this fashion we can continuously monitor one of several possible spin observables that correlate with rotations of the Stokes vector on the Poincaré sphere [14, 44]. For example, if the state of the probe is initially one of linear polarization along $x$, and we choose to measure the probe Faraday rotation then we obtain a (classical) signal proportional to the atomic magnetization $F_z$. If instead we choose to measure the probe ellipticity then we obtain a signal proportional to the atomic alignment, $F_x F_y + F_y F_x$, as discussed in Sec. 2.

Experiments of this type invariably works with atomic ensembles, and one typically assumes the probe field couples identically to every ensemble member. In principle, a measurement of the probe's Stokes vector can then reveal information about the *many-body* state of the collection of spins. For this to be the case, the quantum uncertainty of the measured observable must be greater than the fundamental resolution of the quantum probe [15]. The quantum uncertainty of the observable is often referred to as "projection noise", because repeated strong measurements on identically prepared spins will produce values that fluctuate randomly within the uncertainty distribution, and cause measurement backaction that modify the state accordingly (in the extreme case of a very strong measurement, the backaction is a projection onto an eigenstate of the observable) [45]. The fundamental resolution of the quantum probe is set by shot noise in the polarimeter photodetectors. We show below that for a dilute sample, such as that produced by a typical magneto-optic-trap, shot noise dominates over projection noise and measurement backaction can be neglected on timescales shorter than the photon scattering time. In this situation, if we start with a sample of $N_A$ uncorrelated spins, the mean polarimetry signal is $N_A$ times the signal expected from a single spin, and the probe shot noise alone determines the signal fluctuations. A weak continuous measurement of this type is very useful for single-atom measurement and control, and serves as the basis for, e.g., experimental quantum state reconstruction. Ultimately, the creation and control of many-body states, such as spin-squeezed states of the collective spin, requires ensembles that are optically dense on resonance, an avenue that is currently explored in a number of laboratories [17].

We can quantify the effect of backaction using the theory of completely positive maps [2]. The probe must now be treated as a quantum field with Stokes vector components defined according to Eq. (7). We assume that all of the spins are identically coupled to the same mode of the field, so that the probe couples to the "collective spin" $\mathbf{F} \equiv \sum_i \mathbf{f}^{(i)}$, where for clarity, we henceforth denote the single spins by lower case letters and the collective spin operator with a capital letter. Consider, e.g., a Faraday rotation measurement which correlates the collective atomic magnetization $F_z$ to the field Stokes vector component $S_3$. The coupling is established by the unitary transformation $U_{AL} = \exp[-i\chi F_z S_3]$, where $\chi = \chi_0/(3f)$ is the Faraday rotation angle per unit angular momentum, for a single spin and for a probe detuning much larger than the excited-state hyperfine splitting (see Eqs. (4,9-10)). We further assume that the probe pulse contains a large photon number $N_L$, and is initially prepared with linear polarization in the $x$-direction. For physically reasonable Faraday rotation angles, the 1-component of the Stokes vector is approximately constant, $S_1 \approx \sqrt{N_L/2}$, and the Poincaré sphere is well-approximated by a phase-plane in which the probe is described by a set of canonical coordinates according to the Holstein-Primakov approximation, $X_L = S_2/\sqrt{N_L/2}$, $P_L = S_3/\sqrt{N_L/2}$, $[X_L, P_L] \approx i$. Faraday rotation then corresponds to an $X$-displacement in the "Poincaré plane" proportional to the atomic magnetization, and the corresponding unitary can be written as $U_{AL} = \exp\left[-i(\chi\sqrt{N_L/2}\, F_z)P_L\right]$.

The resolution of the polarimeter follows from the general theory of quantum measurement [2]. In the Holstein-Primakov approximation, the initially *x*-polarized probe pulse is represented by the vacuum state $|0_L\rangle$, which has mean values $\langle X_L\rangle = \langle P_L\rangle = 0$ and vacuum fluctuations $\Delta X_L^2 = \Delta P_L^2 = 1/2$. After the probe is coupled to the spins via the Faraday interaction, the outcomes of measuring $S_2$ are characterized by the eigenvectors $|X_L\rangle$ and Kraus operators $A_{X_L} = \langle X_L|U_{AL}|0_L\rangle$. For a given atomic state $\rho_A$, the probability of observing the value $X_L$ with the polarimeter is $\Pr_{X_L} = \mathrm{Tr}(\rho_A \mathcal{E}_{X_L})$, where the POVM element is given by,

$$\mathcal{E}_{X_L} = A^\dagger_{X_L} A_{X_L} = \left|\langle X_L|e^{-i(\sqrt{N_L/2}\,\chi F_z)P_L}|0\rangle\right|^2 = e^{-\left(X_L - \chi\sqrt{N_L/2}\,F_z\right)^2} = \exp\left[-\frac{\left(F_z - X_L\sqrt{2/\chi^2 N_L}\right)^2}{2(\Delta F_z^2)_{SN}}\right]. \quad (29)$$

The measurement results are Gaussian distributed about a mean $\bar{X}_L = \sqrt{N_L/2}\,\chi\langle F_z\rangle$, with a variance of 1/2 caused by shot noise. Intuitively, the minimum resolvable value of $\langle F_z\rangle$ can be determined by setting the corresponding signal equal to the root-mean-square of the shot noise, which yields $\langle F_z\rangle_{\min} = 1/(\chi\sqrt{N_L})$. More formally, given a measurement value $X_L$, the post-measurement state of the atomic spin ensemble is given by the update rule, $\rho \Rightarrow A_{X_L}\rho A^+_{X_L}/\Pr_{X_L}$. As seen from Eq. (29), the update on the distribution of collective $F_z$ values is a Gaussian filter of width $(\Delta F_z^2)_{SN} = 1/(N_L\chi^2)$, corresponding to the shot noise limited resolution of the polarimeter. One commonly defines the measurement strength in terms of the rate of decrease in the measurement variance [46],

$$\kappa \equiv \frac{1}{(\Delta F_z^2)_{SN}\tau} = \chi^2\frac{N_L}{\tau} = \frac{1}{(3f)^2}\frac{\sigma_0}{A}\gamma_s, \quad (30)$$

proportional to the rate of photon scattering into the probe by a single spin.

The shot-noise limited resolution must now be compared with the quantum fluctuations of the atomic magnetization relative to its expected value. For an ensemble of $N_A$ uncorrelated atoms in a spin coherent state, the projection noise variance is $N_A$ times that of an individual spin, $(\Delta F_z^2)_{PN} = N_A \Delta f_z^2 = N_A f/2$ (the standard quantum limit). Backaction is important when the uncertainty from projection noise dominates over the shot-noise variance, and we can characterize its importance in terms of the key parameter

$$\xi \equiv \frac{(\Delta F_z^2)_{PN}}{(\Delta F_z^2)_{SN}} = \frac{f}{2} N_L N_A \chi^2 = \frac{1}{18f} \rho_{OD} \gamma_s \tau . \qquad (31)$$

Here, the optical density on resonance, $\rho_{OD} = N_A \sigma_0 / A$, and photon scattering rate, $\gamma_s = (I\sigma_0/\hbar\omega)(\Gamma^2/4\Delta^2)$, are defined for a two-level transition with unit oscillator strength, whose optical scattering cross section is $\sigma_0 = 3\lambda^2/2\pi$. Physically, this parameter quantifies the probability of scattering photon into the probe mode. It is this process that leads to entanglement between the spin ensemble and the probe polarization, and the possibility of backaction on the quantum many-body state after a polarization measurement on the probe [12]. Significant backaction thus requires $\xi \gg 1$. In general, there is also some probability for scattering photons into other modes of the electromagnetic field, which leads to decoherence as described by the master equation in Sec. 2.2. This process is characterized by the photon scattering rate $\gamma_s$, and this limits the useful measurement window to times $\tau \sim \gamma_s^{-1}$. As a result, significant backaction can be achieved only for high optical densities on resonance, $\rho_{OD} \gg 1$, much larger than those achieved in a typical magneto-optic trap, but accessible in vapor cells [4] and optical dipole traps [17], or with the enhanced coupling strength available in an optical cavity [21].

For optically thin samples, projection noise is masked by shot noise, and quantum backaction is negligible. In that case, an ensemble of $N_A$ atoms will remain uncorrelated for any time shorter than the coherence time $\tau \sim \gamma_s^{-1}$, and we can assume the collective spin remains in a product state $\rho^{\otimes N_A}$. Continuous measurement of an observable $O$ will then lead to a record that is Gaussian distributed about the expected value. A stochastic realization can be modeled as

$$M(t) = N_A \langle O \rangle_t + \sigma W_t, \qquad (32)$$

where $\langle O \rangle_t = Tr(O\rho(t))$ is the mean signal. The second term represents white noise modeled as a Wiener process, where $W_t$ is a Gaussian random variable of unit variance, and $\sigma^2 = 1/\kappa\tau$ is the shot noise variance.

3.2.2 Quantum-state reconstruction

The combination of weak continuous measurement and quantum control provides a powerful tool for robust and efficient quantum-state reconstruction (QSR). The standard paradigm for QSR involves a series of repeated, strong measurements for each member of an informationally complete set of observables [47]. The procedure is time consuming because the number of real parameters required to specify a general state of $d$-dimensional system scales as $d^2$, and more so because strong backaction erases information about complementary observables and thus requires a freshly prepared copy of the quantum state for each new measurement. Further complexity is added if the measurement apparatus needs to be reconfigured each time a new observable is measured. In contrast, a weak continuous measurement performed simultaneously across a single, large ensemble, combined with a suitable dynamical evolution of the ensemble members, allows us to extract complete information about the state from a continuous measurement record.

In the weak backaction regime, continuous measurement of the observable $O$ leads to a signal of the form given by Eq. (32). The requirement for informational completeness is then equivalent to the requirement of controllability. We can write the time-dependent expectation value in the Heisenberg picture, $\langle O \rangle_t = Tr(\rho_0 O(t))$, where $\rho_0$ is the initial state and $O(t)$ is the time-evolved observable. This illustrates how new information about the initial state becomes available in the measurement record as the system evolves, and shows that complete information is obtained when, in the course of time, $O(t)$ spans the space of Hermitian matrices. In the absence of noise, the measurement record uniquely specifies the state. Shot noise reduces the problem to stochastic estimation, but noise can be averaged out over time if decoherence does not erase the state before complete information is retrieved. For a sufficiently large signal-to-noise ratio, the measurement record from a single ensemble provides a unique fingerprint of the initial state and enables us to perform high fidelity QSR.

A brief description of our QSR algorithm follows. The initial density matrix is decomposed in an arbitrary orthonormal basis of Hermitian matrices $\rho_0 = \sum_\alpha r_\alpha \hat{E}_\alpha$. In general we take the matrix to be unit trace, though in practice population in the subspace of interest can decrease due to loss processes in the experiment. This leaves $d^2 - 1$ basis matrices and real parameters. Discretizing the signal into bins set by the detector bandwidth, the measurement record time-series is

$$M_i = N_A Tr(O(t_i)\rho_0) = N_A \sum_\alpha \tilde{O}_{i\alpha} r_\alpha + \sigma W_i, \tag{33}$$

where $\tilde{O}_{i\alpha} \equiv Tr(O(t_i)\hat{E}_\alpha)$ is a rectangular matrix of real numbers. Because of shot noise, perfect inversion is not possible, but because the statistics are approximately Gaussian, a least-squares-fit (maximum likelihood (ML) estimate) is given by the Moore-Penrose pseudo-inverse,

$$r_\alpha^{ML} = N_A \sum_i (\tilde{O}^T \tilde{O})^{-1}_{\alpha\beta} \tilde{O}^T_{\beta i} M_i. \tag{34}$$

In the absence of noise, the ML estimate will be an exact reconstruction when $\tilde{O}^T \tilde{O}$ is a full-rank square matrix (this defines an informationally complete set of observables). In the presence of noise, the fidelity will be imperfect. In particular, whereas the parameterization we have chosen ensures that the density matrix is Hermitian, it does not ensure that the reconstructed state is positive semidefinite, as required. The closest positive (physical) density matrix to the ML estimate, Eq. (34), can be found efficiently through convex optimization via a linear program [48, 49].

Beyond the finite SNR, the reconstruction algorithm is limited by nonideal dynamics. Fundamentally, probing the spin with an optical field causes decoherence due to photon scattering into modes other than the probe. $O(t)$ thus does not evolve according to a unitary transformation, but instead according to an irreversible map that eventually will map $O$ to the identity, at which time no further information about the state can be retrieved. In addition, inhomogeneities across the sample can interfere with the dynamical control, thereby degrading the accuracy of the intended observables $O(t_i)$ and the information retrieved by the measurement. Successful QSR therefore requires that the dynamical control be fast enough to generate an informationally complete set of observables on a timescale shorter than those set by decoherence and dephasing.

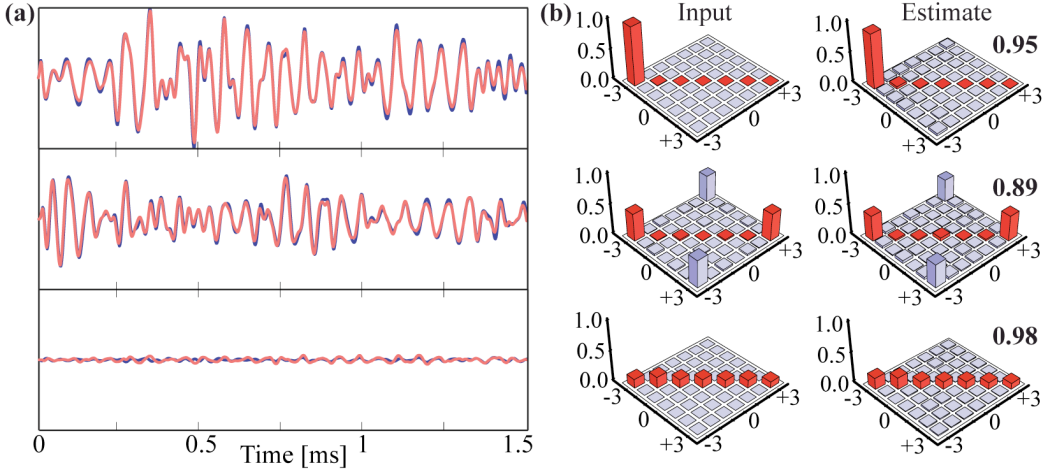

**Fig. 6:** Continuous measurement and QSR. (a) Simulated (dark blue) and observed (light red) measurement signals for test states $|M=3\rangle$ (top), $|\psi_c\rangle = \frac{1}{\sqrt{2}}(|M=3\rangle + i|M=-3\rangle)$ (middle) and a nearly maximally mixed state (bottom). (c) Input and estimated density matrices (absolute values) corresponding to the simulated and observed measurement records. From [25].

The atomic spin/optical system is well suited for the above approach to QSR, and we have implemented it in the laboratory as described in [23,24]. First, we configure our polarimeter to measure ellipticity, i.e. rotation of the probe Stokes vector around the **2**-axis. According to Sec. 2.1, this measurement correlates to the mean atomic moment $\langle O \rangle = \langle F_H F_V + F_V F_H \rangle$. Second, we design a magnetic field waveform that maps this observable onto an informationally complete set in $T \sim 1.5$ ms, using the same general parameter values as for Figs. 3,4 and 5. Figure 6(a) shows three examples of measurement records obtained in the laboratory, compared to the predicted measurement records from a full model simulation based on the master equation (Eq. 17), and including a statistical average over the spatial distribution of probe intensities. The general parameter values are the same as used in Fig. 3. Two points are of note from this figure. Firstly, the simulated measurements are an extremely accurate model of the experimental data, and secondly, each initial state has its own distinct fingerprint in the measurement record. It is these facts in combination that allow high fidelity quantum state construction, as shown by the initially prepared and reconstructed states in Fig. 6(b). It is instructive also to perform QSR based on the measurement record prior to a certain time $t < T$, and then consider the reconstruction fidelity as a function of $t$ as shown in Fig. 7. For short times, long before the dynamics has generated an informationally complete set of observables, the reconstructed state fluctuates randomly and the fidelity is low. After about 300 μs, the generated observables span a sufficient subspace of Hermitian matrices for the extracted information, combined with the positivity constraint on the state, to allow QSR with reasonable fidelity. After 1.5 ms the generated set of observables is near complete, and the QSR fidelity approaches unity even for "difficult" states such as the maximally mixed state.

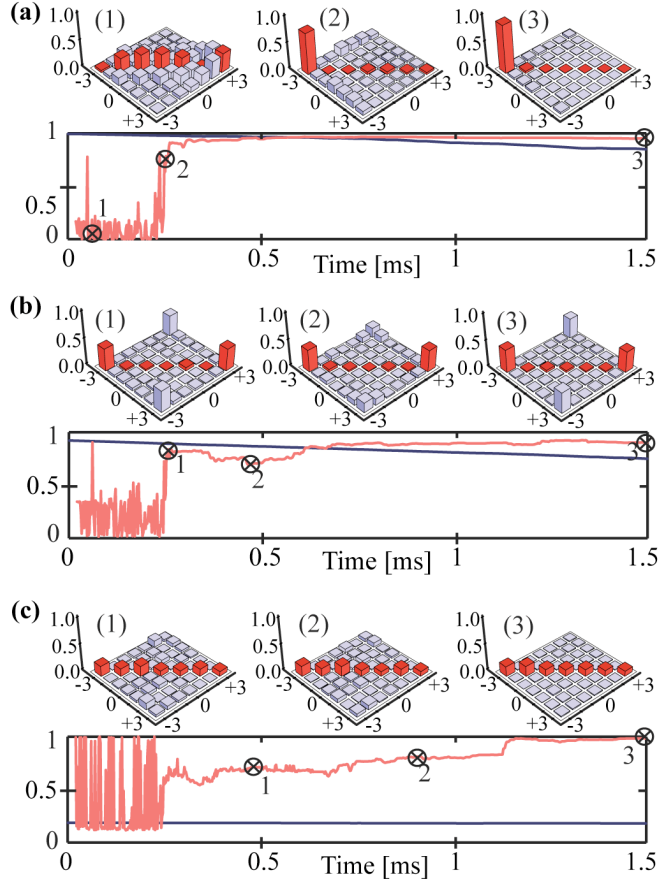

**Fig. 7:** Evolving fidelity, state estimate, and purity during QSR. Plots show the fidelity of the estimate $\rho_{ML}$ (light red) and the largest eigenvalue (dark blue) of the evolving state as a function of elapsed measurement time, for the input states in Fig. 6. Inserts show the estimated density matrices (absolute values only) at a few representative times. From [25].

## 4. Summary and Outlook

In this article we have revisited the quantum interface between an ensemble of alkali atomic spins and an optical probe field as a means to perform quantum-state control and measurement. We have emphasized the dual features of this interaction: the polarization-dependent light-shift acts to drive dynamics of the atomic spin, while the spin-dependent index of refraction acts to affect the probe's polarization dynamics. The former provides an essential ingredient that allows for full quantum control of the atomic dynamics, while the latter allows us to continuously measure ensemble averages of atom-spin observables through polarization spectroscopy.

By introducing an irreducible tensor decomposition, one sees how different moments of the atomic spin distribution are coupled to the components of the field's polarization tensor, parameterized by the Stokes vector. Transmission of the probe through the atomic gas thus leads to rotation of the Stokes vector on the Poincaré sphere, corresponding to birefringence and/or the Faraday effect. Detection in a well-chosen polarization analyzer yields a QND measurement of the corresponding atomic observable.

The irreducible tensor decomposition also gives insight into the field-driven atomic spin dynamics. The rank-1 component gives rise to a Zeeman-like Hamiltonian, with a fictitious magnetic field that is proportional to the ellipticity of the probe polarization. The rank-2 component of the interaction gives rise to a Hamiltonian that is nonlinear in the components of the spin angular momentum operators, and thus to richer dynamics than simple spin rotations. When the excited-state hyperfine splitting is large, the strength of the nonlinearity can be sufficient that it dominates the atomic dynamics over several scattering times. Ultimately, the nonlinear interaction cannot be made arbitrarily large compared to the of rate photon scattering, because the two effects scale with probe parameters in the same way when the detuning is large compared to the excited state hyperfine splitting. In this situation an accurate master equation treatment is essential, and we have included a full derivation in Appendix B, including all optical pumping and light-shift processes.

With theoretical models in hand, we have reviewed the implementations of quantum control and measurement protocols that we have implemented in the laboratory. The electronic-ground-state hyperfine manifolds of $^{133}$Cs, with $F$=3 or 4, span moderately sized Hilbert spaces of dimension $d=2F+1$, on which control is nontrivial. The combination of time-varying magnetic fields and a nonlinear spin rotation arising from the irreducible rank-2 light-shift interaction are sufficient to generate an arbitrary unitary transformation on either of these Hilbert space. Focusing on the $F = 3$ manifold and employing the techniques of optimal control, we found time-dependent control fields that drive a state mapping from a spin-polarized state to an arbitrary superposition of magnetic sublevels. To verify this, one must perform full quantum state tomography. We have designed and implemented a protocol whereby we utilize our capacity to generate an arbitrary unitary map in order to generate an "informationally complete measurement record". Here, the probe field acts both to drive the nonlinear dynamics of the evolving spins that generate new information, and as the meter in a weak measurement of the spin state when analyzed in a polarimeter. The continuous measurement record can then be inverted to determine the initial spin state.

While the tools we have developed so far have been used for high-fidelity control and measurement, they can be substantially improved and extended. Firstly, the nonlinear dynamics of the spins, essential for full controllability of the Hilbert space, arise from a light-shift interaction that is intrinsically tied to photon scattering, and thus has a limited figure of merit for nonlinear evolution vs. decoherence rates. One can achieve the same controllability by introducing resonant microwave and radio-frequency oscillating magnetic fields. We have recently completed an initial theoretical study of this system, and seen that we can achieve full control on the 16 dimensional $6S_{1/2}$ subspace of $^{133}$Cs in

~150 μs, with accessible experimental parameters [50]. Such control must be combined with a continuous measurement protocol to allow for QSR in the entire ground manifold. Secondly, our control objective to date has been state-mapping from an initially known fiducial state. A more challenging and powerful goal is to design control fields that implement a full unitary map on any unknown quantum state. Such maps are of importance for implementing, e.g., unitary logic gates on qudits ($d>2$ subsystems used to store quantum information) [2]. We have developed a constructive protocol to efficiently design these unitary maps that leverages off of the simplicity of state-to-state maps [51]. Implementation of such a protocol will require the design of highly robust, fast, and coherent controls for state mapping.

Finally, we have restricted our attention here to quantum control and measurement of single atomic spins. This leaves out new and interesting physics that plays out in the many-body context. The simplest system to consider in this context is the collective spin consisting of the symmetric subspace of the many-body ensemble. Studies of the quantum interface between the collective spin and polarization spectroscopy have been the focus of a number of research groups. Based of our quantum control perspective, we hope to build upon this work, exploring the production and measurement of highly nonclassical states of the system [52, 53], their dynamics, and their application in quantum information processing [54].

**Acknowledgments:** We are honored to contribute to this special issue in memory of a dear friend and fellow New Mexican of IHD, Krzysztof Wodkiewicz, whose pioneering work in quantum optics has been an inspiration. This article is at its heart pedagogy, and in developing its foundation, we stand on the shoulders of the pioneers that came before us. Krzysztof Wodkiewicz is one of those trailblazers of quantum optics, whose pedagogical approach to theoretical physics set the standards to which we aspire.
We are indebted to Andrew Silberfarb, Greg A. Smith, Souma Chaudhury, Seth Merkel, and Carlos Riofrio for their contributions to the theoretical and experimental developments presented in this article. The research was supported by NSF Grant No.'s PHY-0653599, PHY-0653631, PHY-0555573 and PHY-0555673, by ONR Grant No. N00014-05-1-420, and by IARPA Grant No. DAAD19-13-R-0011.

**Appendix A.**

In this appendix we express the light-shift interaction in terms of its irreducible tensor components using a Cartesian expansion that is most amenable to a basis independent representation. For light detuned between a ground and an excited hyperfine manifold with quantum numbers $F$ and $F'$ respectively, the atomic ac-polarizability tensor is,

$$\vec{\vec{\alpha}}(F,F') = -\frac{P_{g,F}\mathbf{d}P_{e,F'}\mathbf{d}P_{g,F}}{\hbar\Delta_{F'F}}, \tag{A1}$$

where $P_{g,F}, P_{e,F'}$ are projection operators onto the ground/excited subspaces, and $\mathbf{d}$ is the atomic dipole operator. We can extract the characteristic units by expressing $\mathbf{d}$ in terms of its reduced matrix element via the Wigner-Eckart theorem. We define a raising operator

$$\mathbf{D}^{\dagger}_{F'F} \equiv \frac{P_{e,F'}\mathbf{d}P_{g,F}}{\langle nP_{J'}\|d\|nS_J\rangle} = \sum_{q,M,M'} \mathbf{e}_q^* \, o_{JF}^{J'F'} \langle F'M'|FM;1q\rangle \, |F'M'\rangle\langle FM|, \tag{A2a}$$

where

$$o_{JF}^{J'F'} = (-1)^{F'+1+J+I}\sqrt{(2J'+1)(2F+1)}\begin{Bmatrix} F' & I & J' \\ J & 1 & F \end{Bmatrix}. \tag{A2b}$$

are the relative oscillator strengths depending on Wigner 6J symbols and degeneracy factors, determining the branching ratios for spontaneous decay in the hyperfine multiplets by $\Gamma_{J'F'\to JF} = |o_{JF}^{J'F'}|^2 \Gamma_{J'\to J}$. We thus express the polarizability tensor operator as,

$$\alpha_{ij}(F,F') = \alpha_0 \left(\mathbf{e}_i \cdot \mathbf{D}_{FF'}\right)\left(\mathbf{D}^{\dagger}_{F'F} \cdot \mathbf{e}_j\right) \equiv \alpha_0 A_{ij}, \tag{A3}$$

where $\alpha_0 = -\left|\langle nP_{J'}\|d\|nS_J\rangle\right|^2 / \hbar\Delta_{F'F}$ is the characteristic polarizability.

We further employ the Wigner-Eckart theorem to decompose the tensor $A_{ij}$, into its irreducible component. We define the spherical irreducible tensor operators acting in a hyperfine manifold with spin $F$,

$$T_Q^{(K)}(F) = \sum_M |F,M+Q\rangle\langle F,M+Q|FM;KQ\rangle\langle FM| = \frac{\mathcal{Y}_Q^K(\hat{\mathbf{F}})}{\langle F\|\mathcal{Y}^K\|F\rangle}$$

$$= \frac{2^K}{K!}\left[\frac{4\pi(2F+1)(2F-K)!}{(2K+1)(2F+K+1)!}\right]^{1/2} \mathcal{Y}_Q^K(\hat{\mathbf{F}}), \tag{A4}$$

where $\mathcal{Y}_Q^K(\hat{\mathbf{F}})$ are the solid harmonics as a function of the components of the spin operator, and the normalization is chosen so that these tensors have a unit reduced matrix element [55]. The first three tensors, corresponding to the atomic population ($K=0$), orientation ($K=1$), and alignment ($K=2$) are,

$$T_0^{(0)}(F) = 1, \tag{A5a}$$

$$T_Q^{(1)}(F) = \frac{1}{\sqrt{F(F+1)}} F_Q, \tag{A5b}$$

$$T_Q^{(2)}(F) = \frac{\sqrt{6}}{\sqrt{F(F+1)(2F-1)(2F+3)}} \sum_{q,q'} \langle 2Q|1q;1q'\rangle F_q F_{q'}. \tag{A5c}$$

To transform to Cartesian components, express the tensor

$$T_{ij}^{(K)}(F) = \sum_Q (-1)^Q \left(\mathbf{e}_i \cdot \vec{\mathbf{e}}_{-Q}^{(K)} \cdot \mathbf{e}_j\right) T_Q^{(K)}(F), \tag{A6}$$

where the basis dyadics are [3],

$$\vec{\mathbf{e}}_Q^{(K)} = \sum_q \mathbf{e}_q \mathbf{e}_{Q-q} \langle KQ|1Q-q;1q\rangle, \tag{A7a}$$

$$\mathbf{e}_i \cdot \vec{\mathbf{e}}_0^{(0)} \cdot \mathbf{e}_j = -\frac{1}{\sqrt{3}}\delta_{ij}, \quad (-1)^Q \mathbf{e}_i \cdot \vec{\mathbf{e}}_{-Q}^{(1)} \cdot \mathbf{e}_j = \frac{1}{i\sqrt{2}} \varepsilon_{ijk} \mathbf{e}_k \cdot \mathbf{e}_Q^*. \tag{A7b}$$

According to the basis change,

$$T_{ij}^{(0)}(F) = \left(\frac{-1}{\sqrt{3}}\right)\delta_{ij}, \tag{A8a}$$

$$T_{ij}^{(1)}(F) = \left(\frac{-i}{\sqrt{2F(F+1)}}\right)\varepsilon_{ijk}F_k, \tag{A8b}$$

$$T_{ij}^{(2)}(F) = \left(\frac{\sqrt{6}}{\sqrt{F(F+1)(2F-1)(2F+3)}}\right)\left(\frac{1}{2}(F_iF_j + F_jF_i) - \frac{1}{3}\mathbf{F}^2\delta_{ij}\right). \tag{A8c}$$

Using the Wigner-Eckart theorem, we can then express the tensor as

$$A_{ij} = \sum_K \langle F\|A^{(K)}\|F\rangle T_{ij}^{(K)}(F), \tag{A9}$$

where $\langle F\|A^{(K)}\|F\rangle$ is the reduced matrix element. This is related to that of the dipole operators through the 6J symbol,

$$\langle F\|\hat{A}^{(K)}\|F\rangle = \langle F\|(\hat{D}\hat{D}^\dagger)^{(K)}\|F\rangle$$
$$= (-1)^{K+2F}\sqrt{(2F'+1)(2K+1)}\begin{Bmatrix} F & 1 & F' \\ 1 & F & K \end{Bmatrix}\langle F\|\hat{D}\|F'\rangle\langle F'\|\hat{D}^\dagger\|F\rangle. \tag{A10}$$

By definition, $\langle F'\|D^\dagger\|F\rangle = 1$, and from our convention,

$$\langle F\|D\|F'\rangle = (-1)^{F-F'}\sqrt{\frac{2F'+1}{2F+1}}, \tag{A11}$$

leading to a basis independent form of the polarizability tensor,

$$\alpha_{ij}(F,F') = \alpha_0\left[C_{J'FF'}^{(0)}\delta_{ij} + C_{J'FF'}^{(1)}(i\varepsilon_{ijk}F_k) + C_{J'FF'}^{(2)}\left(\frac{1}{2}(F_iF_j + F_jF_i) - \frac{1}{3}\delta_{ij}\mathbf{F}^2\right)\right], \tag{A12}$$

where

$$C^{(0)}_{J'F'F} = (-1)^{3F-F'+1} \frac{1}{\sqrt{3}} \frac{2F'+1}{\sqrt{2F+1}} \begin{Bmatrix} F & 1 & F' \\ 1 & F & 0 \end{Bmatrix} \left|o^{J'F'}_{1/2F}\right|^2, \qquad \text{(A13a)}$$

$$C^{(1)}_{J'F'F} = (-1)^{3F-F'} \sqrt{\frac{3}{2}} \frac{2F'+1}{\sqrt{F(F+1)(2F+1)}} \begin{Bmatrix} F & 1 & F' \\ 1 & F & 1 \end{Bmatrix} \left|o^{J'F'}_{1/2F}\right|^2, \qquad \text{(A13b)}$$

$$C^{(2)}_{J'F'F} = (-1)^{3F-F'} \frac{\sqrt{30}(2F'+1)}{\sqrt{F(F+1)(2F+1)(2F-1)(2F+3)}} \begin{Bmatrix} F & 1 & F' \\ 1 & F & 2 \end{Bmatrix} \left|o^{J'F'}_{1/2F}\right|^2. \qquad \text{(A13c)}$$

**Appendix B**

We derive here the master equation for the ground electronic subspace of an alkali atom, driven off resonantly by an optical field in the low saturation limit, including light-shifts, optical pumping, and hyperfine multiplets. This a generalization of the form used in near resonance polarization-gradient laser cooling, where the field is tuned close to a single hyperfine resonance $F'$ [56]. We begin with the master equation for the $nS_{J=1/2} \to nP_{J'}$ atomic transition with spontaneous emission. In the Lindbald form,

$$\frac{d\rho}{dt} = -\frac{i}{\hbar}\left(H_{eff}\rho - \rho H^\dagger_{eff}\right) + \Gamma \sum_q D_q \rho D^\dagger_q, \qquad \text{(B1)}$$

where the effective non-Hermitian Hamiltonian including hyperfine structure and the light-atom interaction in the rotating wave approximation is

$$H_{eff} = \sum_F E_F P_{g,F} + \sum_{F'}\left(E_{F'} - i\frac{\hbar\Gamma}{2}\right)P_{e,F'} - \frac{\hbar\Omega}{2}\left(e^{-i\omega_L t}\vec{\varepsilon}_L \cdot \mathbf{D}^\dagger + e^{+i\omega_L t}\vec{\varepsilon}^*_L \cdot \mathbf{D}\right). \qquad \text{(B2)}$$

The dimensionless, dipole raising operator, and its conjugate are defined as in Eq. (A2), with $\mathbf{D}^\dagger = \sum_{F,F'} \mathbf{D}^\dagger_{F'F}$. As in Appendix A, we define operators projected between and hyperfine manifolds, $O_{F_1 F_2} \equiv P_{F_1} O P_{F_2} = \sum_{M_1 M_2} \langle F_1 M_1|O|F_2 M_2\rangle |F_1 M_1\rangle\langle F_2 M_2|$. Going to the interaction picture with respect to the free-atom Hamiltonian,

$$H_{eff}^{(I)} = -i\frac{\hbar\Gamma}{2}\sum_{F'} P_{e,F'} - \frac{\hbar\Omega}{2}\sum_{F,F'}\left(e^{-i\Delta_{F'F}t}\vec{\varepsilon}_L \cdot \mathbf{D}_{F'F}^\dagger + e^{+i\Delta_{F'F}t}\vec{\varepsilon}_L^* \cdot \mathbf{D}_{FF'}\right). \tag{B3}$$

Excluding the "feeding term" in the master equation, the dynamics by the effective Hamiltonian is governed by the Weisskoph-Wigner description of nonunitary evolution of a wave function [57]. Defining the ground and excited probability amplitudes of the atomic state, $c_g^{FM} \equiv \langle nS_J; FM | \psi \rangle$, $c_e^{F'M'} \equiv \langle nP_{J'}; F'M' | \psi \rangle$, the Schrödinger equation is

$$\frac{d}{dt}c_e^{F'M'} = -\frac{\Gamma}{2}c_e^{F'M'} + i\frac{\Omega}{2}\sum_{FM} e^{-i\Delta_{F'F}t}\langle F'M'|\varepsilon_L \cdot \mathbf{D}^\dagger|FM\rangle c_g^{FM}, \tag{B4a}$$

$$\frac{d}{dt}c_g^{FM} = i\frac{\Omega}{2}\sum_{F'M'} e^{i\Delta_{F'F}t}\langle FM|\varepsilon_L^* \cdot \mathbf{D}|F'M'\rangle c_e^{F'M'}. \tag{B4b}$$

Adiabatic elimination slaves the rapidly oscillating excited amplitude to the slowly varying ground state [58]. By formal integration,

$$\begin{aligned}c_e^{F'M'} &= i\frac{\Omega}{2}e^{-\frac{\Gamma}{2}t}\sum_{FM}\langle F'M'|\varepsilon_L\cdot\mathbf{D}^\dagger|FM\rangle\int_0^t dt'\, e^{-i\left(\Delta_{F'F}+i\frac{\Gamma}{2}\right)t'} c_g^{FM}(t')\\ &\approx -\frac{\Omega}{2}\sum_{FM} e^{-i\Delta_{F'F}t}\frac{\langle F'M'|\varepsilon_L\cdot\mathbf{D}^\dagger|FM\rangle}{\Delta_{F'F}+i\Gamma/2} c_g^{FM}(t).\end{aligned} \tag{B5}$$

Plugging this into the equation for the ground amplitude evolution,

$$\frac{d}{dt}c_g^{FM} = -\frac{i}{\hbar}\sum_{F',F_1M_1} e^{-i\omega_{FF_1}t}\frac{\Omega^2/4}{\Delta_{F'F_1}+i\Gamma/2}\langle FM|\left(\varepsilon_L^*\cdot\mathbf{D}_{FF'}\right)\left(\varepsilon_L\cdot\mathbf{D}_{F'F_1}^\dagger\right)|F_1M_1\rangle c_g^{F_1M_1}. \tag{B6}$$

Because $\Omega^2/2\Delta_{FF_1} \ll \omega_{F_1F}$ for $F \neq F_1$, Raman-Rabi flopping between different hyperfine manifolds is completely negligible. The effective light-shift interaction is then block-diagonal in the two hyperfine manifolds, with non-Hermitian evolution,

$$\frac{d}{dt}c_g^{FM} = -\frac{i}{\hbar}\sum_{M_1}\langle FM|H_A^{eff}|FM_1\rangle c_g^{FM_1}, \tag{B7}$$

$$H_A^{eff} = -\frac{1}{4}\mathbf{E}_L^* \cdot \vec{\alpha} \cdot \mathbf{E}_L, \quad \vec{\alpha} = -\sum_{F,F'} \frac{P_{g,F}\mathbf{d}P_{e,F'}\mathbf{d}P_{g,F}}{\hbar(\Delta_{F'F} + i\Gamma/2)}. \tag{B8}$$

Finally, to get the complete master equation, we must include the feeding terms for populations and coherences in the ground electronic subspace. In the interaction picture, the matrix-blocks of the ground subspace are fed according to,

$$\left.\frac{d\rho_{F_1M_1;F_2M_2}}{dt}\right|_{feed} = \Gamma \sum_{q,F_1',F_2'} e^{-i(\omega_{F_1'F_2'}-\omega_{F_1F_2})t} \langle F_1M_1|(\mathbf{e}_q^* \cdot \mathbf{D}_{F_1F_1'})\rho_{F_1'F_2'}(\mathbf{e}_q \cdot \mathbf{D}_{F_2'F_2}^\dagger)|F_2M_2\rangle. \tag{B9}$$

The excited state coherences and populations can be expressed in terms of the adiabatically eliminated excited state amplitudes,

$$\langle F_1'M_1'|\rho_{F_1'F_2'}|F_2'M_2'\rangle = (c_e^{F_1'M_1'})(c_e^{F_2'M_2'})^* = e^{i(\omega_{F_1'F_2'}-\omega_{F_\alpha F_\beta})}\frac{\Omega^2}{4}\sum_{F_\alpha F_\beta}\frac{\langle F_1'M_1'|(\varepsilon_L \cdot \mathbf{D}_{F_1'F_\alpha}^\dagger)\rho_{F_\alpha F_\beta}(\varepsilon_L^* \cdot \mathbf{D}_{F_\beta F_2'})|F_2'M_2'\rangle}{(\Delta_{F_1'F_\alpha} + i\Gamma/2)(\Delta_{F_2'F_\beta} - i\Gamma/2)}. \tag{B10}$$

Plugging this back into Eq. (B9),

$$\left.\frac{d\rho_{F_1,F_2}}{dt}\right|_{feed} = \Gamma\sum_{q,F_\alpha,F_\beta} e^{-i(\omega_{F_\alpha F_\beta} - \omega_{F_1F_2})t} W_q^{F_1F_\alpha} \rho_{F_\alpha F_\beta} W_q^{\dagger F_\beta F_2}, \tag{B11}$$

where

$$W_q^{F_bF_a} = \sum_{F'} \frac{\Omega/2}{\Delta_{F'F_a} + i\Gamma/2}(\mathbf{e}_q^* \cdot \mathbf{D}_{F_bF'})(\varepsilon_L \cdot \mathbf{D}_{F'F_a}^\dagger) \tag{B12}$$

are the jump operators for optical pumping between magnetic sublevels according to the Krammers-Heisenberg formula, $\gamma_{F_aM_a \to F_bM_b} = \sum_q |\langle F_bM_b|W_q^{F_bF_a}|F_aM_a\rangle|^2$. The rapidly oscillating terms in Eq. (B11) average quickly to zero unless $\omega_{F_1F_2} = \omega_{F_\alpha F_\beta}$, and can be

completely neglected. This is the secular approximation and leads to an equation in which populations within a given hyperfine manifold are fed by other populations (optical pumping) and coherences are between manifolds fed coherences. It is essential to include the feeding of coherences in order to properly account for indistinguishable scattering processes that do not cause full decoherence [59]. The final form of the master equation in the ground electronic subspace is then,

$$\frac{d\rho}{dt} = -\frac{i}{\hbar}\left(H_A^{eff}\rho - \rho H_A^{eff\dagger}\right) + \Gamma\sum_q\left(\sum_{F,F_1} W_q^{FF_1}\rho_{F_1F_1}W_q^{\dagger F_1F} + \sum_{F_1\neq F_2} W_q^{F_2F_2}\rho_{F_2F_1}W_q^{\dagger F_1F_1}\right). \quad \text{(B13)}$$

This is a trace-preserving completely positive map, accounting for all light shifts, optical pumping, and the effects of decoherence of the atom in the process of photon scattering.